\documentclass[12pt]{article}

\usepackage{amsmath,amsthm, amsfonts, amssymb, amsxtra, amsopn}
\usepackage{pgfplots}
\pgfplotsset{scaled y ticks=false}
\pgfplotsset{compat=1.13}
\usepackage{pgfplotstable}
\usepackage{pgf-pie}
\usepackage{graphicx,grffile}
\usepackage{multirow}
\usepackage{booktabs}
\usepackage{listings}
\usepackage{cmap}
\usepackage{colortbl}
\usepackage{adjustbox}
\usepackage{epsfig}

\usepackage[tableposition=top,font=small,skip=5pt]{caption}
\usepackage{subcaption}
\usepackage{makecell} 

\usepackage[explicit]{titlesec}

\def\urrr#1{\underline{\color{red}#1}}

\PassOptionsToPackage{hyphens}{url}

\usepackage{hyperref}
\hypersetup{colorlinks=true,linkcolor=black,citecolor=black,urlcolor=blue,filecolor=black}
\hypersetup{pdfpagemode=UseNone,pdfstartview=}

\newcommand\dunderline[3][-1pt]{{%
  \sbox0{#3}%
  \ooalign{\copy0\cr\rule[\dimexpr#1-#2\relax]{\wd0}{#2}}}}
\def\uu{\raisebox{1pt}{\dunderline{0.85pt}{\phantom{u}}}}

\usepackage{enumitem}
\setlist[itemize]{noitemsep, topsep=0pt}

\advance\oddsidemargin by -0.35in
\advance\textwidth by 0.7in

\advance\topmargin by -0.4in
\advance\textheight by 0.8in

\def\eref#1{(\ref{#1})}

\long\def\symbolfootnotetext[#1]#2{\begingroup%
\def\thefootnote{\fnsymbol{footnote}}\footnotetext[#1]{#2}\endgroup}

\DeclareMathOperator{\argmin}{argmin}

\def\hh{\phantom{\smash{\displaystyle\frac{1}{24}\Big(\kern1pt\hbox{}}}}
\def\hhh{\phantom{\smash{\displaystyle\frac{1}{4}\Big(\kern1pt\hbox{}}}}

\title{XAI and Android Malware Models}

\author{Maithili Kulkarni\footnotemark[1]\ \ \ 
Mark Stamp\footnotemark[1]\,\,\footnotemark[2]}

\begin{document}

\symbolfootnotetext[1]{Department of Computer Science, San Jose State University}
\symbolfootnotetext[2]{mark.stamp$@$sjsu.edu}

\maketitle

\abstract
Android malware detection based on machine learning (ML) and deep learning (DL) models
is widely used for mobile device security. Such models offer benefits in terms of detection 
accuracy and efficiency, but it is often difficult to understand how such learning models make 
decisions. As a result, these popular malware detection strategies are 
generally treated as black boxes, 
which can result in a lack of trust in the decisions made, as well as making
adversarial attacks more difficult to detect. 
The field of eXplainable Artificial Intelligence (XAI) attempts to shed light on 
such black box models. In this paper, we apply XAI techniques to ML and DL models
that have been trained on a challenging
Android malware classification problem. 
Specifically, the classic ML models considered are 
Support Vector Machines (SVM), Random Forest, 
and $k$-Nearest Neighbors ($k$-NN), while
the DL models we consider are Multi-Layer Perceptrons (MLP) and 
Convolutional Neural Networks (CNN). 
The state-of-the-art XAI techniques that we apply to these
trained models are
Local Interpretable Model-agnostic Explanations (LIME), 
Shapley Additive exPlanations (SHAP), 
PDP plots, 
ELI5, and 
Class Activation Mapping (CAM). 
We obtain global and local explanation results, 
and we discuss the utility of XAI techniques in this problem domain. We also provide 
a literature review of XAI work related to Android malware.

\section{Introduction}

Malicious software, or malware, can appear in various forms, including worms, viruses, adware, 
and ransomware. In recent years, the popularity of smartphones has made them 
targets of malware attacks.

It is not surprising that machine learning (ML) and deep learning (DL) have become dominant
approaches for detecting malware, 
including malware on mobile devices~\cite{stamp2022introduction}. 
Such models can be trained on a variety of 
static and dynamic features~\cite{Anusha,7881800}. 
We elaborate on some of these techniques in Section~\ref{chap:background}.

Although ML and DL provide significant capabilities, 
such techniques are generally treated as black boxes~\cite{google}. This black box aspect
can limit the trust that users are willing to place in such models. Also, from a security perspective, 
black box models may be more susceptible to adversarial attacks, where an attacker attempts 
to modify a model to yield incorrect results. 
Furthermore, when an opaque model fails, it is difficult to identify why the model is 
failing. Thus, there is a need to develop insights into the internal operations of learning models,
especially those that are used for malware detection and classification.

The emerging field of eXplainable Artificial Intelligence (XAI) deals with understanding 
the inner workings of learning models~\cite{molnar2022}. 
XAI generally attempts to explain model outcomes in terms of
the influence of input variable (i.e., features), or by using approximation or surrogate models
whose outcomes are more explainable. 
The goal is to provide a transparent and interpretable view of a model's decision-making process.
In this paper, we focus on XAI in the context of Android malware detection.

We consider XAI for selected classic ML techniques and DL models that have been trained 
on the well-known KronoDroid Android malware dataset. Specifically, the classic ML models
that we consider are Support Vector Machines (SVM), $k$-Nearest Neighbors ($k$-NN), 
and Random Forest. In the DL realm, we consider Multi-Layer Perceptron (MLP) and 
Convolutional Neural Network (CNN) architectures. In general, classic ML techniques are 
relatively interpretable, as ML models are typically based on 
probabilistic, algebraic, or geometric intuition. In contrast, most neural networking models 
are opaque, in the sense that it is non-trivial to understand how they are 
making decisions. In this paper, we aim to provide a comparative study of XAI for 
the selected ML and DL models, within the context of Android malware classification. 

For each trained model, we apply relevant XAI techniques from among the following: 
Local Interpretable Model-Agnostic Explanations (LIME), 
SHapley Additive exPlanations (SHAP), 
PDP, 
and ELI5~\cite{e23010018, Mishra2022, xaisurvey}. 
Additionally, 
we provide a review of recent literature where XAI techniques are applied to models trained
on Android malware. Our literature review can be viewed as an extension of that 
in~\cite{Liu_2022}.

The remainder of this paper is organized as follows. Section~\ref{chap:background} 
covers a range of relevant background topics, including malware detection strategies and
an introduction to the XAI techniques that we employ in our experiments. 
Section~\ref{chap:relwork} gives an overview of related previous work on malware 
classification and provides a literature review of recent XAI work related to 
models trained on Android malware. Section~\ref{chap:results} covers the 
implementation of the various classic ML and DL models used 
in this paper, along with our experiments and results. Finally, 
Section~\ref{chap:conclusion} summarizes our work, 
and we provide a discussion of potential avenues for future work.

\section{Background}\label{chap:background}

In this section, we first give a brief overview of malware, followed by a discussion 
of ML and DL algorithms that are commonly used to classify malware. This section 
also includes detailed background on the state-of-the-art XAI techniques that we
consider in this paper.

\subsection{Malware and Categories }\label{2}

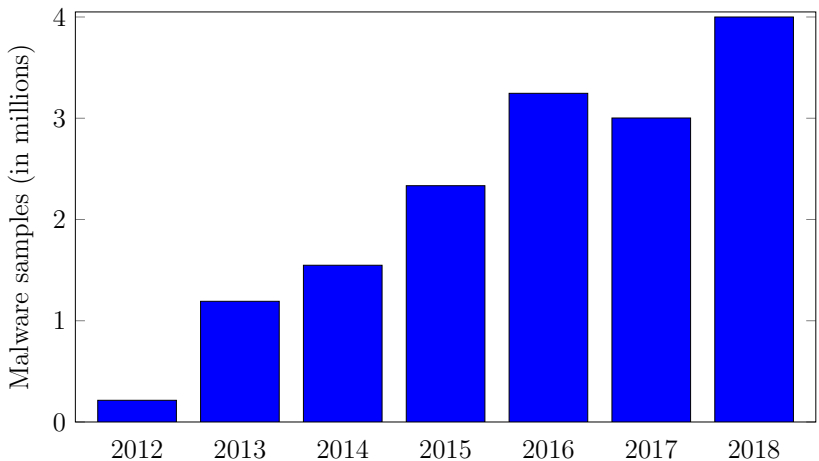
\begin{figure}[!htb]
\centering
\begin{tikzpicture}[scale=0.85, every node/.style={scale=1.0}]
\pgfkeys{/pgf/number format/.cd,1000 sep={}}
\begin{axis}[
        width  = 0.85*\textwidth,
        height = 8.0cm,
        ymin=0,ymax=4.05,
        ytick={0,1,2,3,4},
        major x tick style = transparent,
        ybar=5*\pgflinewidth,
        bar width=35.0pt,
        ylabel = {Malware samples (in millions)},
        symbolic x coords={2012,2013,2014,2015,2016,2017,2018},
        xticklabels={2012,2013,2014,2015,2016,2017,2018},
	y tick label style={
    		/pgf/number format/.cd,
   		fixed,
   		fixed zerofill,
    		precision=0},
        xtick = data,
        x tick label style={
		},
        enlarge x limits=0.1,
]
\addplot [fill=blue,opacity=1.00]%
coordinates {
(2012,0.214327)
(2013,1.192035)
(2014,1.548129)
(2015,2.333777)
(2016,3.246248)
(2017,3.002482)
(2018,4.000000)
};
\end{axis}
\end{tikzpicture}
\caption{Detected Android malware samples}\label{fig:android-mal-trend}
\end{figure}

Malware is the dominant security threat to smartphones~\cite{phacategories}. 
The purpose of writing malware can range from simply a prank 
to an organized criminal activity, information warfare, and espionage. 
Figure~\ref{fig:android-mal-trend} highlights the rapid increase in the volume of 
Android malware samples over the 
years~2012 through~2018~\cite{cyberattacksandroid}.

Malware covers an array of threats, including backdoors, spyware and adware,
Trojan horses, and viruses. We now give a brief overview of these common types of 
malware before moving on to discuss malware detection techniques.

A backdoor, also known as a trapdoor, is built to circumvent 
security checks~\cite{phacategories}. Programmers may create backdoors 
for legitimate reasons when developing their code. 
Cybercriminals seek to exploit their backdoors to delete files, access sensitive data, 
install additional malware, open communication ports for remote access, and so on.

As the name implies, spyware is used to spy on user activities, and can include 
recording the audio of calls on a smartphone, tracking Internet usage, 
recording keystrokes (including passwords), and so on~\cite{phacategories}. 
Adware, on the other hand, often generates fake error messages 
and then asks the user to pay money to fix a non-existent problem. 
Winwebsec is a well-known family of adware~\cite{winwebsecblog}.

Named after the ancient historic plot by Greek invaders to capture Troy, 
a Trojan is a program devised to look harmless, but secretly performs 
a malicious task. Trojan apps that send premium SMS messages in the background 
are a typical example~\cite{phacategories}. Zeus (aka Zbot) is a well-known Trojan family, 
and it has been widely used for nearly two decades for crimes including bank fraud 
and money laundering~\cite{zeusblog}.

A virus is the most common type of malware. True to its name, this
malware replicates by infecting executable programs. The infected programs can
further propagate the virus during their execution, or a virus might propagate 
through external devices, software, or email. Like a biological virus, a computer virus
might exhibit metamorphism, in the sense of changing its form when infecting other
systems~\cite{stamp2011infosec}. Metamorphism is an effective means of evading
classic virus detection techniques, such as signature scanning.

\subsection{Learning Models for Malware Detection}

There is a constant arms race between malware writers and antivirus developers.
Over the past two decades, ML and DL techniques have come to the fore
in the fields of malware detection, classification, and analysis. In this section, 
we introduce the ML and DL techniques that we consider in this paper,
where the underlying problem is to classify Android malware samples.

\subsubsection{Classic Machine Learning}

Support Vector Machines (SVM)~\cite{stamp2022introduction} are popular supervised machine 
learning models. SVMs attempt to separate classes using hyperplanes. A 
nonlinear kernel can be used to map training data into a higher-dimensional space and 
thereby enhance the separability. 

In machine learning, a Random Forest consists of an ensemble of 
decision trees, with voting among the component trees used to
determine the classification~\cite{ibmblog}. More trees can mean better accuracy
and generalizability, but care must be taken not to overfit the data.

As the name suggests, in $k$-Nearest Neighbors ($k$-NN), samples are classified based 
on the~$k$ nearest samples in the training set. There is no explicit training required 
for $k$-NN, and hence the algorithm is often referred to as a ``lazy learner''. 
However, scoring 
calculation can be relatively expensive. The technique 
is highly sensitive to local structure and, in particular, for small values of $k$, 
overfitting is common~\cite{stamp2022introduction}.

\subsubsection{Deep Learning}

Artificial Neural Networks (ANNs) are mathematical models that are inspired by neurons 
in the brain. Multi-layer Perceptrons (MLP) are the simplest useful neural networking 
architecture, and hence they are sometimes referred to simply as ANNs. 
MLPs are feed-forward networks that generalize basic perceptrons to allow for nonlinear 
decision boundaries. This is analogous to the way that 
nonlinear SVMs generalizes linear SVMs. As with most DL architectures,
MLPs are trained using backpropagation~\cite{stamp2022introduction}.

Convolutional Neural Networks (CNNs) are a specialized type of neural network that focus 
on local structure, making them ideal for image analysis. CNNs are composed of an input layer, 
convolution layers, and pooling layers, along with a fully-connected output layer (or layers)
that produce a vector of class scores. 
The first convolutional layer in a CNN extracts various intuitive features from 
the input. Subsequent convolutional layers extract ever more abstract features from the
previous layer.

\subsection{Overview of Explainable AI}

The applications of artificial intelligence in the security domain introduces several challenges. 
For example, adversarial attacks on such systems are a concern. By employing 
eXplainable Artificial Intelligence (XAI) techniques
to understand how a model works, we have a better chance of detecting such attacks. 
Additionally, XAI analysis, may enable us to perform feature reduction,
based on feature importance, which can speed up detection. 
XAI can shed light on black box models by uncovering relationships between dependent and 
independent variables, thereby increasing user trust, which is especially important
in security-related applications.

Next, we briefly consider XAI techniques from various perspectives. Specifically, we
discuss interpretability and explanations from the perspectives of 
ante-hoc versus post-hoc, model-agnostic versus model-specific,
and local versus global. We then consider the level of 
interpretability---high, medium, or low---provided by XAI techniques

Models that are inherently easy to understand are said to be ante-hoc interpretable. 
For example, linear models and classic Hidden Markov Models (HMM) fall into the 
ante-hoc interpretable category. A model is post-hoc model interpretable if we need
to apply explicit interpretation methods after the model is trained. Of course, post-hoc techniques 
can also be used on intrinsically interpretable models after they are trained~\cite{xaisurvey}. 

Some XAI techniques are model agnostic, in the sense that they can be
applied to any type of machine learning algorithm. On the other hand, 
some XAI techniques are model-specific. Of the XAI techniques that we consider, 
LIME, SHAP, PDP plots, and ELI5 are all model-agnostic techniques, 
while CAM is specific to CNNs. According to~\cite{molnar2022},
model-specific techniques may, in general, be more 
informative than model-agnostic explanation techniques.

Local interpretable techniques help us understand how and why the model makes 
a certain classification for a specific sample, or for a group of samples~\cite{mathworksinterpret}. 
Locally, models can often be viewed as linear or monotonic in some features.
Global techniques deal with interpreting a model as a whole, 
taking a holistic view of features into account. For example, LIME only deals with
local interpretability, while SHAP can be used for both local and global explanations.

Models consisting of linear functions are highly interpretable. For example,
linear SVM models are highly interpretable. For this reason, some XAI 
techniques, such as LIME, use linear functions as local approximations.

Models with nonlinear monotonic functions are in the class of medium interpretable models. 
Nonlinear functions are those in which input data is modeled using a function with a nonlinear 
combination of the model parameters. For example, an SVM trained with the 
(nonlinear) RBF kernel is a medium interpretable model.

Machine learning models with nonlinear and non-monotonic functions fall into the low interpretability 
category. Most DL models are in this category, and hence they are
inherently difficult to interpret. CAM is model-specific technique that is applicable 
to CNNs, which are in the low interpretability category.

It has been suggested that there may be an inverse correlation between 
model performance and inherent interpretability~\cite{gunning2019xai,mlinterpretintro}.
However, there are XAI techniques that are useful for models in the low interpretability category;
for example, CAM is useful for interpreting CNN models, as mentioned above.

\subsection{XAI Techniques}\label{2.6}

Before moving on to discuss our experiments, we first introduce the explainability techniques 
that we consider. We use feature ranking to analyze our linear SVM and Random Forest 
models, and for other models, we use the XAI techniques of 
LIME, ELI5, CAM, and SHAP (including PDP plots).

\subsubsection{SVM and Random Forest Interpretations}

Linear SVMs are inherently interpretable models,
in the sense that we can determine the relative importance of features 
based on the model weights, assuming that the features have
been properly normalized. In the \texttt{sklearn} Python 
library, it is easy to obtain feature weights for the linear SVM
kernel using the \texttt{coef\uu method}~\cite{scikitlearn}.
Similarly, we can obtain feature rankings from Random Forest models.
Non-linear SVMs, as well as the other ML and DL techniques that we consider, 
are not highly interpretable.

\subsubsection{LIME}

Local Interpretable Model-agnostic Explanations (LIME) is based on local surrogate 
interpretable models, and is used to explain individual predictions of black box machine learning models~\cite{limegit}. LIME generates a new dataset consisting of perturbed samples 
and the corresponding predictions of the black box model. Based on this new dataset, LIME 
then trains a simple interpretable model which is weighted by the proximity of the perturbed 
instances to the sample of interest. This interpretable model provides a good local approximation 
to the original machine learning model.

According to~\cite{Rothman},
the explanation provided by LIME of sample~$x$, denoted~$E(x)$,
can be expressed as
$$
    E(x) =  \underset{g \in G}{\argmin}\big(L(M, g, \pi_x) + \Omega(g)\big)
$$
where~$L$ measures the inaccuracy introduced by approximating the original
model~$M$ with the simplified model~$g$ in a perturbed neighborhood defined by~$\pi_x$.
By default,~$g$ is a sparse linear model, but decision trees can also be used.  
Here, $\Omega(g)$ is a measure of model complexity and acts
as a penalty term, since we want a simple approximation.
Note that the minimum is over the family~$G$ of possible explanations.

Obtaining LIME explanations consists of the following steps. 
\begin{enumerate}
\item Choose a dataset.
\item Train a black box model on the dataset.
\item Generate new data samples by perturbing existing samples and
	weight the new dataset samples according to their proximity 
	to the sample of interest.
\item Train a weighted, interpretable model on this new dataset.
\item Explain the prediction by interpreting the local model.
\end{enumerate}

\subsubsection{ELI5}

The name ELI5 is derived from the saying, ``Explain it Like I'm 5''.
ELI5 can be used to generate global explanations of a black-box model. 
The concept behind ELI5 is simply based on permuting the values of 
individual features---in turn, the values of each feature are shuffled,
and model results are tabulated after each such shuffle. The worse the 
classification results after a given feature is shuffled, the more that the model 
depends on that feature~\cite{eli5}.

\subsubsection{Grad-CAM}

The technique of Gradient-weighted Class Activation Map (Grad-CAM) is used
to analyze CNNs. Grad-CAM assists 
in understanding which parts of an image a convolutional layer weights most
to determine a given classification~\cite{molnar2022}. That is, 
Grad-CAM is a class-based localization technique for CNN interpretability. 

Grad-CAM uses gradient information 
flowing into the last convolutional layer of a CNN to obtain a localization map of the 
important regions in the image, and thereby determines the importance of each pixel 
of the input image for the specified class.
This resulting gradient weighted activation map can be overlayed on
the original input image to visualize which parts of the input the 
CNN associates highly with a given output class.

\subsubsection{SHAP and PDP Plots}\label{2.6.3} 


SHapley Additive exPlanations (SHAP) is a popular XAI technique
based on Shapley values.
In 1951, Lloyd Shapley developed a technique to determine the contribution of 
each player in a multi-player game setting, and in 2012, he won the Nobel Prize 
in economics for his work. In Shapley's approach, player contributions are 
determined by Shapley values, which have a number of desirable theoretical 
properties. More recently, Shapley values have been applied to 
XAI~\cite{lundberg2017unified}, with features in place of game-theoretic players.

In SHAP, we first compute a Shapley value
for each sample and each feature, as discussed below in some detail.
A Shapley value measures the
contribution of a specified feature to the classification of a given sample.
If we arrange the Shapley values into a matrix with the rows indexed 
by the samples and the columns indexed by the features, then the row 
corresponding to a sample can provide an explanation for the classification
of the sample. For example, the largest value in a row corresponds to 
the feature that has the most influence on the classification of the 
corresponding sample. Similarly, explanations of the overall model 
can be determined by analyzing the Shapley values in the entire matrix.

Several types of graphs and plots can be generated based on Shapley values.
Before discussing such graphs, we first provide more details on the
computation of Shapley values.

Suppose that~$X$ represents a
feature vector of length~$n$ of the form~$X=(f_1,f_2,\ldots,f_n)$,
where each~$f_j$ is the value of a specific feature. Further, 
suppose that we have a model~$M$ that for each such~$X$ produces
a real-valued result, $M(X)$. For example, $M(X)$ could be 
the classification of~$X$ as determined by the model~$M$,
or it could be a probability generated by the model.
For any subset~$S$ of the features~$\{f_1,f_2,\ldots,f_n\}$, 
let~$M_S$ be a model corresponding to~$M$, 
but trained only on the features in the subset~$S$.
Then~$M_S(X)$ is the real-valued result obtained for 
sample~$X$, under the restricted model~$M_S$.

For a given sample~$X$, we compute~$n$ Shapley values, with each 
value corresponding to one of the~$n$ features. We denote the Shapley
value for sample~$X$, corresponding to feature~$f_i$, as~$\Phi_i(X)$.
The Shapley value is defined as
\begin{equation}\label{eq:shapVal_a}
	\Phi_i(X) = \frac{1}{n} \sum_{S_i} \Biggl[\big(M_{\{S_i\cup {f_i}\}}(X) - M_{S_i}(X)\big)
  	\Big/ {{n-1} \choose {|S_i|}}\Biggr]
\end{equation}
where~$S_i$ denotes a subset of the~$n-1$ features~$\{f_1,\ldots,f_{i-1},f_{i+1},\ldots,f_n\}$,
and the sum is over all such subsets (including the empty set, with~$M_{\varnothing}(X)$
defined to be~0). 
Note that the Shapley value 
computation consists of comparing the behavior of pairs of models applied to the sample~$X$:
One models of each pair includes the feature~$f_i$, while the other omits~$f_i$, with the other
features unchanged. These pairwise computations are averaged over the number of subsets of
a given size. 
The~$1/n$ term in~\eref{eq:shapVal_a} normalizes the result based on the number of features.
For example, suppose that we have four features
with~$X=(f_1,f_2,f_3,f_4)$, and that we are computing the Shapley value~$\Phi_3(X)$.
Then from equation~\eref{eq:shapVal_a}, we have
\begin{align}\label{eq:shap3}
\begin{split}
	\Phi_3(X) &= \frac{1}{4}
		\Big(\big(M_{\{f_3\}}(X)-M_{\varnothing}(X)\big) \\[-0.5ex]
		&\hhh + \big(M_{\{f_1,f_3\}}(X)-M_{\{f_1\}}(X)\big)/3 \\
		&\hhh + \big(M_{\{f_2,f_3\}}(X)-M_{\{f_2\}}(X)\big)/3 \\
		&\hhh + \big(M_{\{f_3,f_4\}}(X)-M_{\{f_4\}}(X)\big)/3 \\
		&\hhh + \big(M_{\{f_1,f_2,f_3\}}(X)-M_{\{f_1,f_2\}}(X)\big)/3 \\
		&\hhh + \big(M_{\{f_1,f_3,f_4\}}(X)-M_{\{f_1,f_4\}}(X)\big)/3 \\
		&\hhh + \big(M_{\{f_2,f_3,f_4\}}(X)-M_{\{f_2,f_4\}}(X)\big)/3 \\[-0.85ex]
		&\hhh + \big(M_{\{f_1,f_2,f_3,f_4\}}(X)-M_{\{f_1,f_2,f_4\}}(X)\big)\Big)
\end{split}
\end{align}

Shapley values can also be computed by 
considering all~$n!$ orderings of the features. In this formulation, for each
permutation, we again sum the differences of pairs of a models, where
one is trained on all features up to and including~$f_i$, 
with the model trained on all features up to~$f_i$, but not
including~$f_i$. We now discuss this approach in more detail.

For any permutation~$P$ of the features, let~$P_i$ be the initial part of the
permutation before~$f_i$ appears. Then we can rewrite equation~\eref{eq:shapVal_a} as
\begin{equation}\label{eq:shapVal_b}
	\Phi_i(X) = \frac{1}{n!} \sum_{P} 
	\big(
		M_{P_i\cup \{f_i\}}(X) - M_{P_i}(X)
	\big)
\end{equation}
where the sum is over all~$n!$ permutations~$P$ of the~$n$ features~$\{f_1,f_2,\ldots,f_n\}$.

Using equation~\eref{eq:shapVal_b},
the example in equation~\eref{eq:shap3} can be written as
\begingroup
\addtolength{\jot}{-0.25em}
\begin{align}\label{eq:shap3b}
\begin{split}
	\Phi_3(X) &= \frac{1}{24}
		\Big(
		\big(M_{\{\urrr{f_1,f_2,f_3},f_4\}}(X)-M_{\{\urrr{f_1,f_2},f_3,f_4\}}(X)\big) \\
		&\hh + \big(M_{\{\urrr{f_1,f_2,f_4,f_3}\}}(X)-M_{\{\urrr{f_1,f_2,f_4},f_3\}}(X)\big) \\
		&\hh + \big(M_{\{\urrr{f_1,f_3},f_2,f_4\}}(X)-M_{\{\urrr{f_1},f_3,f_2,f_4\}}(X)\big) \\
		&\hh + \big(M_{\{\urrr{f_1,f_3},f_4,f_2\}}(X)-M_{\{\urrr{f_1},f_3,f_4,f_2\}}(X)\big) \\
		&\hh + \big(M_{\{\urrr{f_1,f_4,f_2,f_3}\}}(X)-M_{\{\urrr{f_1,f_4,f_2},f_3\}}(X)\big) \\
		&\hh \hspace*{0.05in}\vdots  \hspace*{0.02in}
					\phantom{ +\big(M_{\{\urrr{f_4,f_1,f_2,f_3}\}}(X)}\vdots \\
		&\hh + \big(M_{\{\urrr{f_4,f_1,f_3},f_2\}}(X)-M_{\{\urrr{f_4,f_1},f_3,f_2\}}(X)\big) \\
		&\hh + \big(M_{\{\urrr{f_4,f_2,f_1,f_3}\}}(X)-M_{\{\urrr{f_4,f_2,f_1},f_3\}}(X)\big) \\
		&\hh + \big(M_{\{\urrr{f_4,f_2,f_3},f_1\}}(X)-M_{\{\urrr{f_4,f_2},f_3,f_1\}}(X)\big) \\
		&\hh + \big(M_{\{\urrr{f_4,f_3},f_1,f_2\}}(X)-M_{\{\urrr{f_4},f_3,f_1,f_2\}}(X)\big) \\[-0.5ex]
		&\hh + \big(M_{\{\urrr{f_4,f_3},f_2,f_1\}}(X)-M_{\{\urrr{f_4},f_3,f_2,f_1\}}(X)\big) 
		\Big)
\end{split}
\end{align}
\endgroup
where, for clarity, we have listed the entirety of each permutations, 
with the underlined red parts
representing the subscripts that appear in~\eref{eq:shap3}. Note that if there is
no underlined part of a permutation, the model is~$M_{\varnothing}$.

From the formula in~\eref{eq:shapVal_b}---and the example in~\eref{eq:shap3b}---we 
can clearly see how the Shapley value~$\Phi_i(X)$ measures the contribution of 
feature~$f_i$ to the classification of sample~$X$. Specifically,
a model is trained on a set of features that includes~$f_i$, 
and the classification of~$X$ by that model is compared to 
that obtained using the same features, except 
that~$f_i$ is removed. Such comparisons are computed for all
permutations, and the results are averaged. 
Rearranging terms, we see that the Shapley value
is the difference between the expected outcome when feature~$f_i$ is included
in the model, and the expected outcome when feature~$f_i$ is not included.

In many cases,
training models for all permutations would be prohibitively 
costly, even for just one Shapley value. Sampling methods
are used, and some of the properties of Shapley values
can also play a role in making the problem computationally tractable.

As alluded to in the previous paragraph,
Shapley values satisfy several useful and interesting properties.
For our purposes the most relevant properties are the following.
\begin{itemize}
\item Efficiency --- The sum of the Shapley values for~$X$ is equal to
the value that the model trained on all features produces for~$X$.
That is,
$$
	M(X) = \sum_{i=1}^n \Phi_i(X) 
$$ 
\item Symmetry --- If~$M_{S\,\cup\{f_i\}}(X) = M_{S\,\cup\{f_j\}}(X)$ for all
feature subsets~$S$ that include neither~$f_i$ nor~$f_j$, 
then~$\Phi_i(X)=\Phi_j(X)$.
\item Linearity --- The Shapley values are linear with respect to samples, that 
is, $\alpha\Phi_i(X)=\Phi_i(\alpha X)$ and~$\Phi_i(X)+\Phi_i(Y)=\Phi_i(X+Y)$.
\item Null --- The Shapley value of a null feature is~0, where a null 
feature, by definition, satisfies~$M_{S\,\cup \{f_i\}}(X)=M_{S}(X)$ 
for all~$S$ that do not include~$f_i$. 
\end{itemize}
The linear property implies that for a Random Forest, we can compute the
Shapley values of each component decision tree and then combine the results
to obtain the Shapley value for the overall model~\cite[Section~9.5]{molnar2022}. 
A similar statement holds for
boosting methods, and hence for both Random Forest and boosting models,
computing Shapley values is computationally feasible.



Partial Dependence Plots (PDP) use Shapley values
to visualize the marginal effect of a predictor variable 
on the predictive variable by plotting the average model outcome at different 
levels of the predictor variable~\cite{molnar2022}. This gives the average effect 
that a predictor variable has on the predictive variable.
These values are plotted on a chart which provides evidence 
of the direction in which a variable affects the outcome.

\section{Related Work}\label{chap:relwork}

XAI is a very active field, although research into its application 
in the malware domain is more limited. In this section, we survey 
previous research that involves applications of XAI to 
malware classification and detection.

Manthena, et al.~\cite{2023xaimanthena} consider XAI in the context
of malware detection. 
The goal of this research is to determine how malware influences the behavior of 
virtual machines (VMs) in a cloud environment. Three different variants of 
SHAP are applied (KernelSHAP, TreeSHAP, and DeepSHAP), while the learning
techniques considered are linear SVM, nonlinear SVM (with RBF kernel), Random Forest, 
a specific feed-forward neural network, and CNN, all of which are trained on a
malware dataset. The researchers use the SHAP interpretations 
to implement feature reductions.

Yan, et al.~\cite{xaisurvey} consider ante-hoc and post-hoc explanation in detail. 
They evaluate these techniques based on six metrics (accuracy, sparsity, completeness, 
stability, efficiency, and fidelity), and conclude that
Layerwise Relevance Propagation (LRP) is the most efficient XAI technique.
The authors also list open issues, including the potential tradeoff between accuracy 
and explanability. 

Charmet, et al.~\cite{2022xaiarticle} provide a comparative study of XAI for different 
cybersecurity tasks with the goal of determining which explanation methods could be 
efficiently used for each of the following: Improved trust (in the sense of 
increased transparency), improved classifier performance, and to explain errors in the models. 
They also show that XAI methods involving heatmaps and saliency maps can be 
easily compromised. 

Ullah, et al.~\cite{2022xais22186766} conduct XAI experiments in
the context of Android malware detection, based on both traditional features and 
greyscale image data. They consider pre-trained Bidirectional Encoder Representations 
from Transformers (BERT) models, which rely on transfer learning. 
LIME and SHAP are used to determine the effect of each feature on the 
overall accuracy of the model.

Liu, et al.~\cite{liu2022explainable} also use XAI approaches to explore learning models
in the realm of malware detection. 
They consider LIME and SHAP, and the research primarily focuses
on understanding the impact of temporal inconsistencies in the training data
with respect to the performance of ML-based malware detection approaches.

Kinkead, et al.~\cite{2021KINKEAD2021959} consider the problem of explaining 
predictions of Android malware classification models. They consider CNN models,
and they use the LIME for their XAI analysis. The authors claim that
their work provides additional trust and confidence in their CNN model.

Severi, et al.~\cite{2021Severi} develop a model-agnostic methodology based on 
SHAP to examine the vulnerability of classifiers to adversarial attack. The research 
is based on static and dynamic analysis of diverse datasets, including
Portable Executable (PE) files and Android applications. 
High-contributing features are selected using SHAP and 
attacks are conducted against a variety of learning models,
including Random Forest, SVM, and a Deep Neural Network (DNN). 
These researchers claim that their explanation-guided attack method is more robust,
as compared to alternative approaches.

Fan, et al.~\cite{2020DBLP} provide guidelines to assess the 
quality, stability, and robustness 
of XAI approaches. They experiment with LIME, Anchor, 
Local Rule-based Explanations (LORE), SHAP, and LEMNA and
consider several learning \hbox{techniques} (MLP, Random Forest, SVM).
They claim that in the domain of Android malware detection, 
inconsistencies in results from different XAI techniques makes it 
difficult to trust the explanations.

Warnecke, et al.~\cite{2020quant} provide general recommendations related
to the application of explanation methods for deep learning techniques 
in the security domain. 
A variety of XAI methods are considered, and the authors find that the
Integrated Gradients and LRP methods are most effective, 
according to their specified criteria.

In Table~\ref{tab:summaryprevwork}, we summarize the papers mentioned in
this section, as well as a few other relevant research papers. We note that
these cited papers are relatively recent, with all having been
published between~2016 and~2023.

\begin{table}[!htb]
\caption{Selected previous work}\label{tab:summaryprevwork}
\centering
\adjustbox{scale=0.85}{
\begin{tabular}{c|cc} \midrule\midrule
	Authors & Dataset & XAI techniques \\ \midrule 
        Charmet, et al.~\cite{2022xaiarticle} & ---  & Survey paper \\ 
	Chen, et al.~\cite{2020denas} & AndroZoo & LEMNA \\ 
        Fan, et al.~\cite{2020DBLP} & Multiple sources & LORE, SHAP, others \\ 
	Feichtner and Gruber~\cite{2019upa} & PlayDrone & LIME \\ 
	Iadarola, et al.~\cite{2021idl} & Android Malware Dataset  & Grad-CAM \\ 
	Kinkead, et al.~\cite{2021KINKEAD2021959} & Drebin & LIME \\ 
	Liu, et al.~\cite{liu2022explainable} & AndroZoo  & LIME, SHAP \\ 
	Manthena, et al.~\cite{2023xaimanthena} & VirusTotal & SHAP \\ 
   	Severi, et al.~\cite{2021Severi} & Grad-CAM & SHAP \\ 
	Ullah, et al.~\cite{2022xais22186766} & CICMalDroid 2020 & LIME, SHAP \\ 
	Warnecke, et al.~\cite{2020quant} & Drebin, Genome & LRP, LIME, SHAP \\ 
        Wu, et al.~\cite{wu2020android} & Drebin+ & X{\sc Mal} \\ 
        Yan, et al.~\cite{xaisurvey} & ---  & Survey paper \\ 
        Yang, et al.~\cite{2021cade} & Drebin &  Distance-based \\ 
	\midrule\midrule
\end{tabular}
}
\end{table}

\begin{figure}[!htb]
\centering
\begin{tikzpicture}[scale=0.9]
\begin{axis}[ 
		   width=0.85\textwidth,
		   height=0.4\textwidth,
	 	   x tick label style={scale=0.85,
   		 	/pgf/number format/.cd,
			/pgf/number format/1000 sep={},
   			fixed,
   			fixed zerofill,
    			precision=0,
			/tikz/.cd
		   },
		   x label style={scale=0.85},
	 	   y tick label style={scale=0.85,
    		 	/pgf/number format/.cd,
   			fixed,
   			fixed zerofill,
    			precision=4,
			/tikz/.cd
		    },
		   y label style={scale=0.85},
                    ymin=0.0,ymax=0.0008,
                    symbolic x coords={2016(4),
                    	2017,2017(2),2017(3),2017(4),
                    	2018,2018(2),2018(3),2018(4),
			2019,2019(2),2019(3),2019(4),
			2020,2020(2),2020(3),2020(4),
			2021,2021(2),2021(3),2021(4),
			2022,2022(2),2022(3),2022(4),
			2023,2023(2)},
                    xtick={
                    	2017,
                    	2018,
			2019,
			2020,
			2021,
			2022,
			2023},
                    ytick={0.0000,0.0002,0.0004,0.0006,0.0008},
                    ylabel={Proportion of research papers},
                    enlarge x limits=0.01,
                    legend pos=north west,
                    legend style={nodes={scale=0.85},
                         }
                    ] 
\addplot[color=red,very thick,mark=none] coordinates { 
(2016(4),0.000)
(2017,0.000)
(2017(2),0.000)
(2017(3),0.000)
(2017(4),0.000)
(2018,0.000)
(2018(2),0.000)
(2018(3),0.000)
(2018(4),0.000)
(2019,0.0001)
(2019(2),0.0001)
(2019(3),0.0001)
(2019(4),0.0002)
(2020,0.000)
(2020(2),0.0002)
(2020(3),0.0001)
(2020(4),0.0002)
(2021,0.0001)
(2021(2),0.0002)
(2021(3),0.0001)
(2021(4),0.0002)
(2022,0.0001)
(2022(2),0.0004)
(2022(3),0.0001)
(2022(4),0.0003)
(2023,0.0002)
(2023(2),0.0003)
};
\addlegendentry{CAM}
\addplot[color=blue,very thick,mark=none] coordinates { 
(2016(4),0.000)
(2017,0.000)
(2017(2),0.000)
(2017(3),0.0001)
(2017(4),0.000)
(2018,0.000)
(2018(2),0.000)
(2018(3),0.000)
(2018(4),0.000)
(2019,0.000)
(2019(2),0.0001)
(2019(3),0.0001)
(2019(4),0.0002)
(2020,0.0002)
(2020(2),0.0001)
(2020(3),0.0002)
(2020(4),0.0002)
(2021,0.0002)
(2021(2),0.0004)
(2021(3),0.0005)
(2021(4),0.0003)
(2022,0.0004)
(2022(2),0.0002)
(2022(3),0.0003)
(2022(4),0.0004)
(2023,0.0006)
(2023(2),0.0004)
};
\addlegendentry{SHAP}
\addplot[color=green,very thick,mark=none] coordinates { 
(2016(4),0.000)
(2017,0.0001)
(2017(2),0.000)
(2017(3),0.000)
(2017(4),0.000)
(2018,0.000)
(2018(2),0.0001)
(2018(3),0.000)
(2018(4),0.0002)
(2019,0.0001)
(2019(2),0.000)
(2019(3),0.0001)
(2019(4),0.0002)
(2020,0.0003)
(2020(2),0.0002)
(2020(3),0.0002)
(2020(4),0.0004)
(2021,0.0004)
(2021(2),0.0003)
(2021(3),0.0004)
(2021(4),0.0003)
(2022,0.0002)
(2022(2),0.0002)
(2022(3),0.0003)
(2022(4),0.0002)
(2023,0.0003)
(2023(2),0.0002)
};
\addlegendentry{LIME}
\end{axis}
\end{tikzpicture}
\caption{XAI research papers}\label{fig:xaitoolusage}
\end{figure}
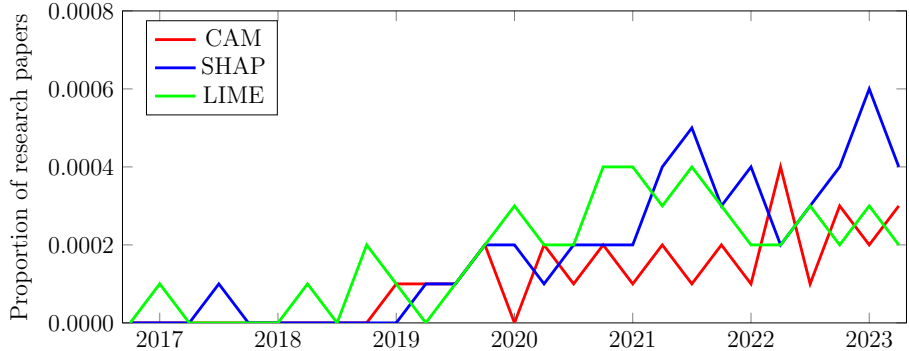

Amongst the XAI techniques considered in this paper, 
SHAP appears to be
the most widely studied by the research community, 
followed by CAM and LIME. 
The graph in Figure~\ref{fig:xaitoolusage} 
charts the appearance of these three XAI technique
in research papers over recent years~\cite{pwctoolusage}. 

We note in passing that the number of relevant studies focusing on evaluating XAI 
in the malware domain is relatively small. 
Further, there is currently no accepted standard method or 
criteria for selecting or evaluating XAI methods for malware-related 
problems, and hence a general recommendation as to which XAI method 
or methods will work well in the Android malware domain is unavailable. 
Thus, more research is needed in this area to determine the practical utility 
of XAI techniques for real-world Android malware problems.

\section{Experiments and Results}\label{chap:results}

In this section, we consider a range of XAI experiments. 
But first, we discuss our dataset and
implementation. 

\subsection{Dataset and Implementation}\label{4.1}

We use the KronoDroid dataset~\cite{kronodroid} for all of the experiments
reported in this paper. 
This dataset includes labeled data from~240 malware families, 
with~78,137 total samples, of which~41,382 are malware and~36,755 
are benign Android apps. For each sample, 289 dynamic features (based on 
system calls) and~200 static features (e.g., \texttt{permissions}) are provided.
Each malware family contains a number of samples collected over an 
extended period of time. Samples belonging to a malware family generally 
have similar characteristics and share a common code base. 

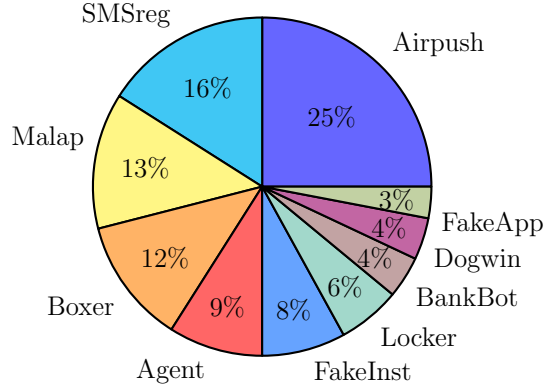
\begin{figure}[!htb]
\centering
\begin{tikzpicture}[scale=0.75, every node/.style={scale=0.85}]
\pie{25/Airpush,
16/SMSreg,
13/Malap,
12/Boxer,
9/Agent,
8/FakeInst,
6/Locker,
4/BankBot,
4/Dogwin,
3/FakeApp}
\end{tikzpicture}
\caption{Top 10 malware families}\label{fig:top10pie-chart}
\end{figure}

To ensure a significant number of samples per family, we restrict our attention
to the top~10 malware families in the KronoDroid dataset.
These top~10 malware families have a total of~31,046 samples,
with the percentage of samples per family illustrated in 
the pie graph in Figure~\ref{fig:top10pie-chart}.


All classic machine algorithms experiments are performed on a single host machine,
while deep learning experiments are performed using the GPU on this same machine. 
All experiments in this research have been executed on the computer 
specified in Table~\ref{tab:compresources}.

\begin{table}[!htb]
\caption{Computing resources used in experiments\label{tab:compresources}}
\centering 
\adjustbox{scale=0.85}{
\begin{tabular}{c|c}\midrule\midrule
Computing resource & Details\\ \midrule
Computer & Dell XPS 13 \\
Processor & Intel Core i5-7200U CPU @ 2.50 Ghz, 2.70 Ghz \\
RAM & 8.0 GB \\
Operating System & Windows 10 Enterprise 64-bit \\ \midrule\midrule
\end{tabular}}
\end{table}

We cleaned the data to remove irrelevant features. The cleaned 
dataset contains~468 features per sample. All features are standardized
using a standard scaler.

In our experiments, we use accuracy and F1-score to measure the performance of each classifier. 
Accuracy is defined as the total number of correct predictions over the number of samples tested. 
The F1-score is a weighted average of \mbox{precision} and recall, and it is computed as
$$ 
  \mbox{F1} = {\tt 2} \times \frac{(\mbox{Precision} \times \mbox{Recall})}{(
  	\mbox{Precision} + \mbox{Recall})}
$$ 
where 
$$ 
  \mbox{Precision} = \frac{\mbox{True Positives}}{(\mbox{True Positives} + \mbox{False Positives})}
$$ 
and
$$ 
  \mbox{Recall} = \frac{\mbox{True Positives}}{(\mbox{True Positives} + \mbox{False Negatives})}
$$
As with accuracy, F1 scores fall between~0 and~1, 
with~1 being the best possible.

As discussed above, the primary goal of this research is to explore
the utility of XAI techniques in the 
Android malware domain. Towards this end, we generate explanations and obtain interpretations 
for SVM (both linear and non-linear), Random Forest, $k$-NN, MLP, and CNN. 
We consider a wide range of XAI experiments, from generating ante-hoc explanations 
based a model's inherent interpretability to post-hoc explanations. We 
generate post-hoc explanations using LIME, SHAP, ELI5, and PDP Plots.
For CNNs, we use the model-specific technique of CAM. 
The package \texttt{scikit-learn} has been 
employed for most of the experiments, with the exception being that 
the \texttt{Tensorflow} and \texttt{Keras} libraries are utilized for CNNs. 
In all cases, we perform stratified 5-fold cross-validation.

A summary of the main hyperparameters for our various models follows. 
\begin{itemize}
\item \textbf{SVM} --- We perform preliminary tests to determine the best kernel for our 
nonlinear SVM, with the result being the Gaussian radial basis function (RBF).
\item \textbf{Random Forest} --- Based on small-scale experiments, we found that 
using~$\mbox{\texttt{n\uu estimator}} = 100$ and otherwise
using the hyperparameter defaults in scikit-learn yielded the best results.
\item \textbf{$k$-NN} --- Again, based on small-scale experiments, 
we selected~$k$ = 5 for all $k$-NN experiments reported in this paper.
\item \textbf{MLP} --- We use a deep architecture with~300 hidden layers, 
rectified linear unit (ReLu) activation functions, and a learning rate of $\alpha = 0.0001$.
\item \textbf{CNN} --- We use max pooling for our CNN model. We experimented
with various hyperparameters and found that an initial number of convolution filters 
set to~32, a filter size~$2\times 2$, and a dropout rate of~$0.25$ yielded the best results.
\end{itemize}

\subsection{Performance of Learning Models}

For the experiments in this section, we use an~80:20 stratified training:testing split.
As mentioned above, all models are trained using only the top~10 malware families 
in the KronoDroid dataset. The results of our experiments are shown 
in Table~\ref{tab:f1score}. We observe that Random Forest performs best,
while MLP is second best. In addition, all models perform reasonably well, with
the accuracy of the worst-performing model being within~4\%\ of that of Random Forest.

\begin{table}[!htb]
	\caption{Performance of ML and DL models}\label{tab:f1score}
	\centering
	\adjustbox{scale=0.85}{
		\begin{tabular}{c|cccccc} \midrule\midrule
		Model & Accuracy & Precision & Recall & F1\\
			\midrule
			Linear SVM       & 0.9180 & 0.9194 & 0.8719 & 0.8917 \\
            		RBF-SVM         & 0.8917 & 0.8937 & 0.8917 & 0.8898 \\
            		Random Forest & \fbox{0.9322} & \fbox{0.9318} & \fbox{0.9322} & \fbox{0.9314} \\
			$k$-NN             & 0.9061 & 0.9052 & 0.9061 & 0.9054 \\
			MLP                  & 0.9209 & 0.9206 & 0.9209 & 0.9207 \\
			CNN                 & 0.9076 & 0.9089 & 0.8976 & 0.9091 \\ \midrule\midrule
		\end{tabular}
		}
\end{table}

\subsection{XAI Results}

In this section, we apply the explainability techniques in Section~\ref{2.6} 
to our models, and we discuss the results. Note that three versions of SHAP
are considered here: For SVM models we use KernalSHAP, for Random Forest
we use TreeSHAP, and for MLP we use DeepSHAP.

\subsubsection{Linear SVM and Random Forest Feature Importance}

\begin{figure}[!htb]
\centering
  \includegraphics[scale=0.5]{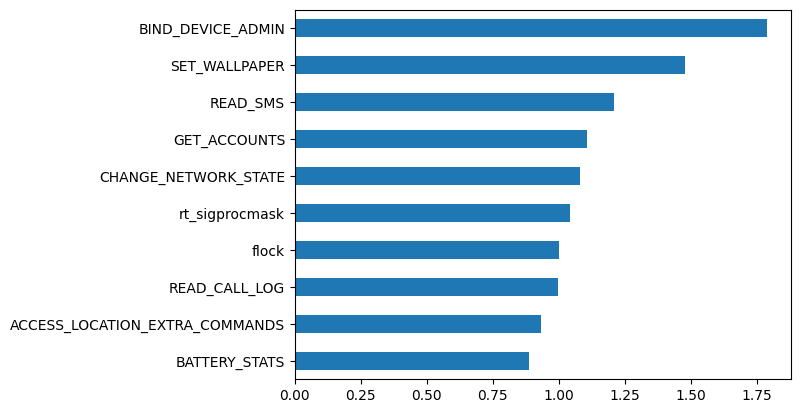}
  \caption{Feature importance from linear SVM}\label{fig:featureimp_SVM}
\end{figure}
\begin{figure}[!htb]
\centering
  \includegraphics[scale=0.5]{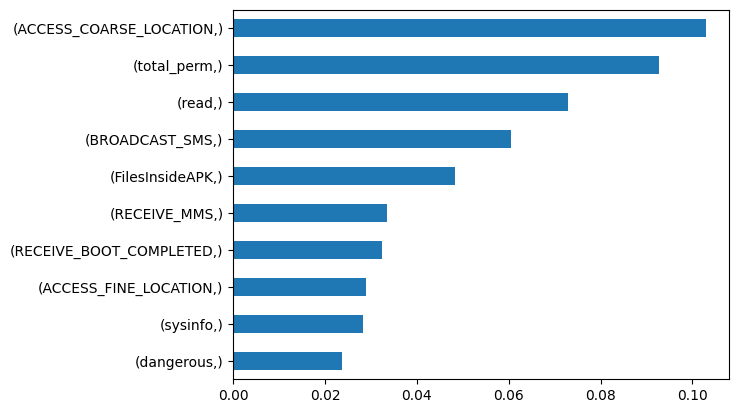}
  \caption{Feature importance from Random Forest}\label{fig:featureimp_RF}
\end{figure}

We calculate feature importance by
extracting the feature weights from the linear SVM and Random Forrest 
models. Figures~\ref{fig:featureimp_SVM} and~\ref{fig:featureimp_RF} show the 
top~10 most important features for our linear SVM and Random Forrest
models, respectively. We observe that
\texttt{BLIND\uu DEVICE\uu ADMIN}, \texttt{SET\uu WALLPAPER}, 
and \texttt{READ\uu SMS} are the main drivers of model predictions for the
linear SVM, while for Random Forrest, \texttt{ACCESS\uu COARSE\uu LOCATION}, 
\texttt{total\uu perm}, and \texttt{read} contribute the most. We find that the 
feature importance results on the train and test sets 
are consistent for both models, which indicates that they are not overfitting 
on the KronoDroid dataset. Extracting such feature coefficients is not possible for a 
nonlinear SVM kernel. 

\subsubsection{ELI5 Feature Importance}\label{4.7.1}

Recall that ELI5 is a permutation-based technique
that measures the change in model 
error after the values of a single feature have been shuffled. 
We use the ELI5 library in Python to calculate permutation 
importance~\cite{ELI5_lib}. 

Table~\ref{tab:rf-eli5} shows the permutation importance for our Random Forest model. 
The values at the top of the ELI5 output are the most important features in our model, 
while those at the bottom matter the least. The first number in each row indicates how much 
the model performance decreased with random shuffling, using the same performance metric 
as the original model---in this case, we use mean squared error (MSE).
The number after the~$\pm$ 
measures how performance varied over the reshuffling, in terms of the range of values. 
For example, shuffling the data of the \texttt{SEND\uu SMS} feature
caused the Random Forest MSE to vary by~0.0010.
By this measure, the top three features are \texttt{SEND\uu SMS}, 
\texttt{RECEIVE\uu BOOT\uu COMPLETED}, 
and \texttt{TimesSubmitted}. 

\begin{table}[!htb]
\caption{ELI5 feature importance for Random Forest}\label{tab:rf-eli5}
\centering
\adjustbox{scale=0.825}{
\begin{tabular}{c|c}
\midrule\midrule
Weight & Feature \\ \midrule
$0.0033 \pm 0.0010$ & \texttt{SEND\uu SMS} \\
$0.0032 \pm 0.0003$ & \texttt{RECEIVE\uu BOOT\uu COMPLETED} \\
$0.0021 \pm 0.0020$ & \texttt{TimesSubmitted} \\
$0.0015 \pm 0.0011$ & \texttt{GET\uu ACCOUNTS} \\
$0.0014 \pm 0.0016$ & \texttt{FilesInsideAPK} \\
$0.0012 \pm 0.0006$ & \texttt{GET\uu TASKS} \\
$0.0011 \pm 0.0017$ & \texttt{UFileSize} \\
$0.0011 \pm 0.0005$ & \texttt{READ\uu EXTERNAL\uu STORAGE} \\
$0.0010 \pm 0.0002$ & \texttt{READ\uu PHONE\uu STATE} \\
$0.0008 \pm 0.0006$ & \texttt{dangerous} \\
$0.0008 \pm 0.0013$ & \texttt{signature} \\
$0.0008 \pm 0.0002$ & \texttt{SYSTEM\uu ALERT\uu WINDOW} \\
$0.0007 \pm 0.0011$ & \texttt{mprotect} \\
$0.0006 \pm 0.0005$ & \texttt{WRITE\uu SECURE\uu SETTINGS} \\
$0.0005 \pm 0.0009$ & \texttt{sysinfo} \\
$0.0005 \pm 0.0004$ & \texttt{CHANGE\uu CONFIGURATIONS} \\
$0.0005 \pm 0.0009$ & \texttt{fsync} \\
$0.0005 \pm 0.0012$ & \texttt{prctl} \\
$0.0004 \pm 0.0004$ & \texttt{READ\uu LOGS} \\
$0.0004 \pm 0.0008$ & \texttt{fchmod} \\
$\vdots$ & $\vdots$ \\
 (448 more) & (448 more) \\
\midrule\midrule
\end{tabular}
}
\end{table}

We note that only four of the top~10 features listed 
in the bar graph in Figure~\ref{fig:featureimp_RF}
appear among the top~20 features
in Table~\ref{tab:rf-eli5}. This points to the issue of
inconsistency between XAI analysis techniques.

\subsubsection{LIME Interpretations}\label{sect:LIME_int}

LIME provides a list of the importance of each feature in model prediction relative to  
a specified sample. Recall that LIME relies on a simplified local model for feature ranking.

KronoDroid dataset consists of tabular data,
so we define a tabular explainer object in LIME. 
The trained model, features used in training, and labels of target classes
serve as inputs, and the results are based on 
the test data.

Figures~\ref{fig:lime_correct}(a) through~(d) in the appendix 
show the LIME explanations for the RBF-SVM, $k$-NN, Random Forest, and MLP 
models, respectively, based on the first sample of the test dataset for each model. 
All models correctly classify this first sample of test data with high confidence
as Locker ransomware. 
The left side of the LIME explanation shows the probability 
with which the sample is classified as ransomware---the pink color indicates 
that the contribution is towards the ransomware family, while the purple color 
indicates that the contribution is towards Malap family. We observe that
these figures show that the RBF-SVM, $k$-NN, Random Forest, 
and MLP models classify this specific sample as 
ransomware with probabilities of~0.82, 1.0, 1.0, and~1.0, respectively. 

The LIME output in Figure~\ref{fig:lime_correct}(a) shows the classification result 
for the top two highest probability classes for this specific sample.
In the middle of the figure, there is a list of rules that gives the reason why this sample 
belongs to the class ransomware, and it identifies and lists the features 
that contribute most to the prediction, in order of importance. On the right side 
of the figure, there is a table---pink values 
are the reason for the final prediction, while green values are the reasons that 
do not support the prediction outcome. In this case, \texttt{SEND\uu SMS}
points strongly towards a ransomware classification,
while there are four features that 
are against the ransomware classification, but only weakly so.



Figures~\ref{fig:lime_incorrect}(a) through~(d) in the appendix 
show the LIME explanations for RBF-SVM, $k$-NN, Random Forest, and MLP models,
respectively, for a sample in the Malap family that is misclassified as BankBot
by all of these models. The (mis)classification probabilities are~0.93, 0.79, 0.57, and~1.0
for the RBF-SVM, $k$-NN, Random Forest, and MLP models. respectively.
In the figures, orange values are the reason for the final prediction, 
and green color values are those that do not support the predicted outcome. 
Figure~\ref{fig:lime_incorrect}(a), for example, shows
that the feature \texttt{SEND\uu SMS} contributes to
a Bankbot and a ransomware classifications, both of which are incorrect,
but since there are fewer negative factors for Bankbot, it is the selected classification. 
Interestingly, Random Forest is the only model that gives any significant 
weight to the possibility of this sample being in the (correct) Malap family, 
but only with a probability of~0.17. 



We observe that the LIME interpretations for the 
RBF-SVM and MLP models are the most similar pair in Figure~\ref{fig:lime_correct}
and, arguably, also in Figure~\ref{fig:lime_incorrect}.
This is not surprising, as nonlinear SVMs and MLPs are closely related models,
in the sense that an MLP can be viewed as an SVM-like model,
where the equivalent of the kernel function is learned~\cite{stamp2022introduction}.
Based on the LIME interpretations in these figures, Random Forest
appears to be the most different from the other three models.
It is somewhat surprising that the $k$-NN and Random Forest 
results are not more similar, as those techniques are both 
neighborhood-based techniques~\cite{stamp2022introduction}.

\subsubsection{Grad-CAM Interpretation}

For this experiment, we represent the input array as an image. 
To generate the images, we first order the~468 features
from highest to lowest importance, as determined by the Random Forest model.
For each sample, we put these ordered feature values into a~$22\times 22$ array
(with~0 padding for the final~16 elements), which we then interpret
as a grayscale image for our CNN model.

We use \texttt{iNNvestigate} library to generate Grad-CAM output on 
our CNN model output.
The method \texttt{create\uu analyzer} of \texttt{iNNvestigate} 
determines the components of the input 
that correspond to the output. It then determines the importance of an input 
pixel based on how much a change in the pixel affects the output. 

\begin{figure}[!htb]
\centering
\begin{tabular}{cc}
  \raisebox{-0.2cm}{\includegraphics[width=57.5mm]{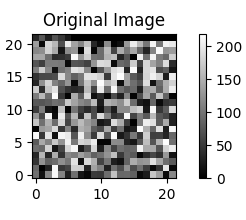}}
&
  \raisebox{0.1cm}{\includegraphics[width=60mm]{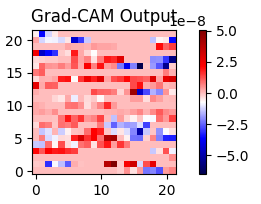}}
\\
\adjustbox{scale=0.95}{(a) Original image}
&
\adjustbox{scale=0.95}{(b) Grad-CAM}
\end{tabular}
\caption{Grad-CAM example}\label{fig:gradcam}
\end{figure}

We analyze an image from the test dataset with the \texttt{gradient} function, 
which gives the gradient of the output neuron with respect to the input. 
Figure~\ref{fig:gradcam}(a) shows the sample test image reshaped 
as $22\times 22$ grayscale image as discussed above. 
Figure~\ref{fig:gradcam}(b) shows the Grad-CAM output for the prediction 
made by the CNN model for this sample. We can visually verify which pixels 
(equivalently, features) in the input image 
the CNN is emphasizing when making its classification. For example, the Grad-CAM
image shows a dark red pixel in row~1, column~12, indicating that the corresponding
feature is one of the most important to the CNN classification
of this particular sample.

We observe that the feature importance determined by Grad-CAM is much different
from that of the Random Forest model. This follows, since the features in the 
original image are ordered from highest to lowest importance, 
according to the Random Forest model weights, 
yet there is only a slight bias towards more important
features in the lower region of the Grad-CAM image.
We conclude that the Random Forest and CNN models 
appear to be using much different criteria to make 
their classification decisions.

\subsubsection{SHAP Interpretations and PDPs}

We use \texttt{KernelSHAP} to generate explanations for our SVM and $k$-NN models, 
\texttt{DeepSHAP} for our MLP,
and \texttt{TreeSHAP} for the Random Forest model. 
It is well-known that \texttt{KernelSHAP} and \texttt{DeepSHAP}
are much more costly to compute, as compared to
\texttt{TreeSHAP}~\cite{jyang}. 

\begin{figure}[!htb]
\centering
\includegraphics[width=90mm]{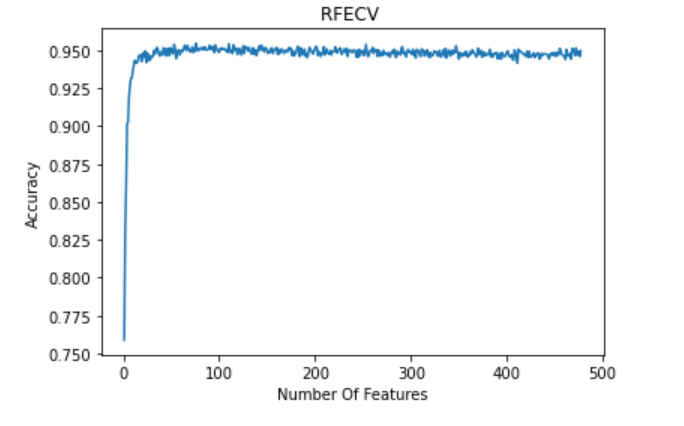}
\caption{Random Forest RFE accuracy}
\label{fig:rfe-accuracy-noFeatures}
\end{figure}

Due to the high computational cost
we use Recursive Feature Elimination (RFE) based on Random Forest models
to determine which features to 
sample. The graph in Figure~\ref{fig:rfe-accuracy-noFeatures} 
shows that the accuracy of the Random Forest model does not improve, 
provided that at least the top~10 features are selected.
Hence, we select these top~10 features to sample for
each of the models under consideration.

For our experiments, \texttt{TreeSHAP} only required about~53 
seconds to complete execution on a dataset of size~41,382,
while \texttt{KernelSHAP} required about~1 hour 
for a comparable experiment. We found that
\texttt{DeepSHAP} was comparable in runtime to \texttt{KernelSHAP}.
For comparison, for the LIME experiments discussed in Section~\ref{sect:LIME_int},
the execution time was on the order of~30 seconds.

Using global model interpretation techniques, 
we can see how our model behaves in general. 
Toward this end, we generate two SHAP global model interpretation plots, 
specifically, a SHAP variable importance plot and a SHAP dependence plot. 

\begin{figure}[!htb]
\advance\tabcolsep by -15pt
\centering
\begin{tabular}{cc}
  \includegraphics[scale=0.32]{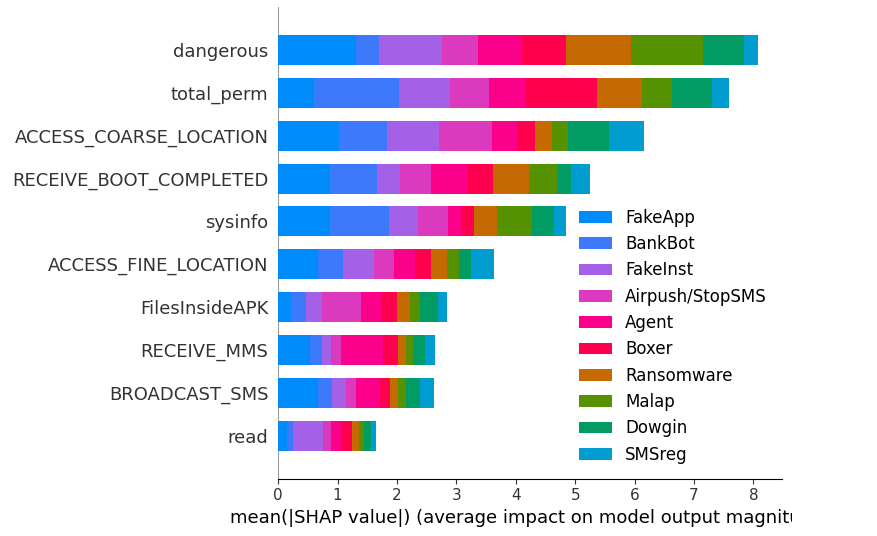}
  &
  \includegraphics[scale=0.32]{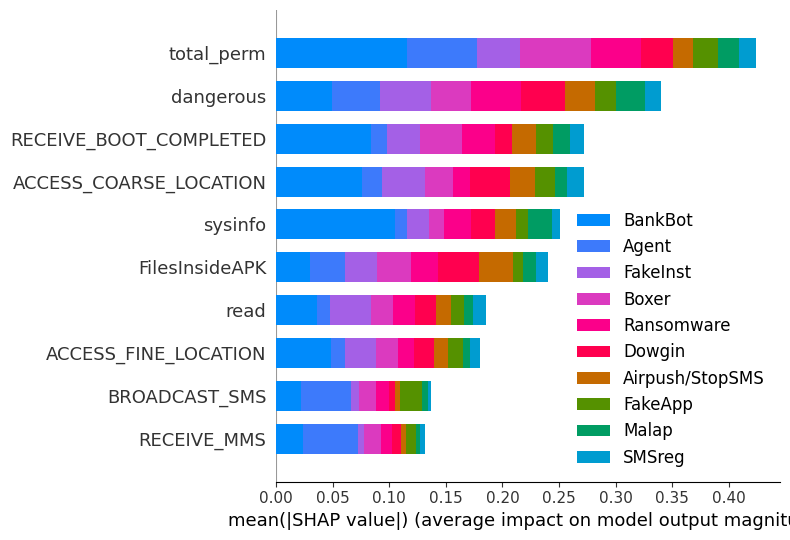}
  \\
  \adjustbox{scale=0.95}{\ \ \ (a) RBF-SVM}
  &
  \adjustbox{scale=0.95}{\ \ \ (b) Random Forest}
  \\ \\
  \includegraphics[scale=0.32]{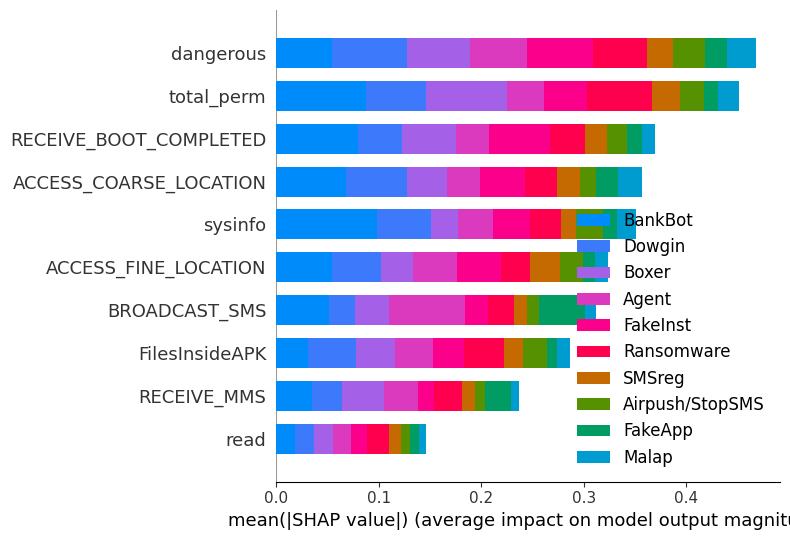}
  &
  \includegraphics[scale=0.32]{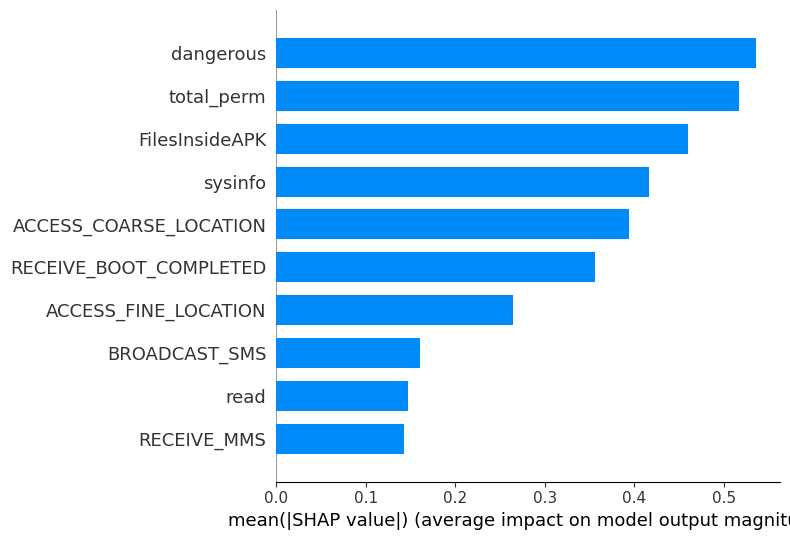}
  \\
  \adjustbox{scale=0.95}{\ \ \ (c) $k$-NN}
  &
  \adjustbox{scale=0.95}{\ \ \ (d) MLP}
  \end{tabular}
  \caption{Variable importance plots}\label{fig:svmshapglobal}
\end{figure}

We use \texttt{shap.summary\uu plot} function with \texttt{plot\uu type} set to \texttt{bar} 
to generate the variable importance plots. 
Figures~\ref{fig:svmshapglobal}(a) through~(d)
provide these SHAP global explanations 
for the RBF-SVM, Random Forest, $k$-NN, and MLP models, respectively. 
In these plots, the $x$-axis denotes the average impact on the model output
(i.e., the mean SHAP value across all relevant samples) 
of the specified variable. 
It is interesting to note that 
the top two ranking features for all of the models
are \texttt{dangerous} and \texttt{total\uu perm}. 
These graphs enable us to easily compare the relative 
contribution of the listed features for each model.

\begin{figure}[!htb]
\centering
\includegraphics[width=90mm]{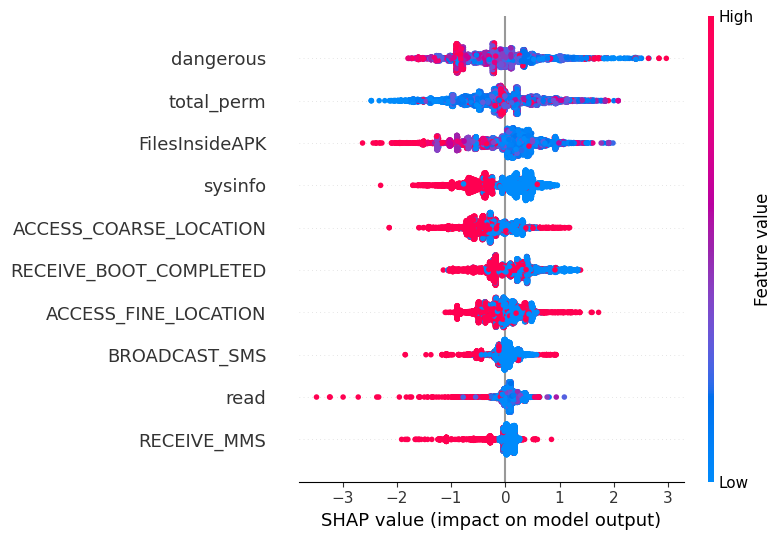}
\caption{MLP global interpretation value plot}
  \label{fig:mlpshapglobal2}
\end{figure}

The SHAP values appear in the form of a beeswarm plot 
in Figure~\ref{fig:mlpshapglobal2}.
The function \texttt{shap.summary\uu plot} was used to produce this plot.
Here, the~$x$-axis indicates the Shapley value,
while the~$y$-axis lists the~10 features under consideration.
Shapley values corresponding to a given feature are plotted for 
all samples in the test set, with the thickness of the ``swarm''
representing the density of points. The color-coding represents the
raw value of the feature, with blue indicating a low value and 
red corresponding to a high value.
Thus, we obtain insight into the relationship of raw features
and their predictive strength via the Shapley values.

From Figure~\ref{fig:mlpshapglobal2}, we make the following observations.
\begin{enumerate}
\item The plot lists the features in descending order of importance which, of course,
matches the results in Figure~\ref{fig:svmshapglobal}(d).
\item For most of the features, raw values that are low
are more predictive than high values, with this being 
especially clear for the \texttt{sysinfo} and \texttt{FilesInsideAPK} features.
\item Curiously, the two highest ranked features behave somewhat differently than
the other features. Specifically, the raw high-low values 
of the feature \texttt{total\uu perm} appears to
have little correlation to the corresponding Shapley values and, to a
somewhat lesser extent, 
this also appears to be the case for the \texttt{dangerous} feature.
\end{enumerate}

Partial Dependence Plots (PDP) show the average manner in which 
machine-learned response functions changes, based on the values of two input 
variables of interest, while averaging out the effects of all other input variables. 
PDP plots enhance our understanding of a model by showing interactions 
between variables and dependent variables in complex models. 
PDP plots can also enhance trust, provided that observed relationships 
conform to domain knowledge expectations.

\begin{figure}[!htb]
  \centering
  \begin{tabular}{cc}
  \includegraphics[scale=0.35]{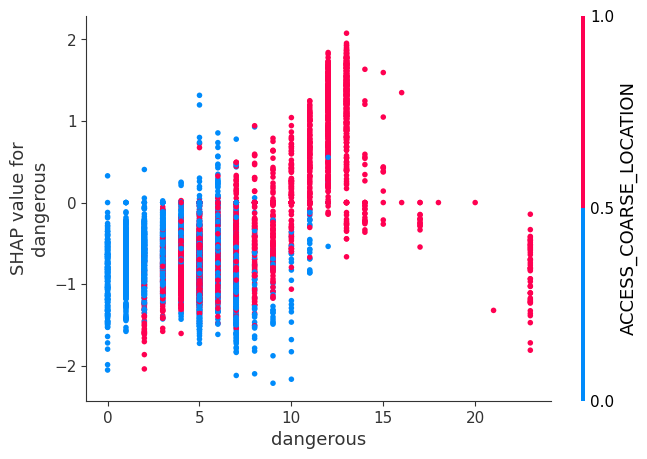}
  &
  \includegraphics[scale=0.35]{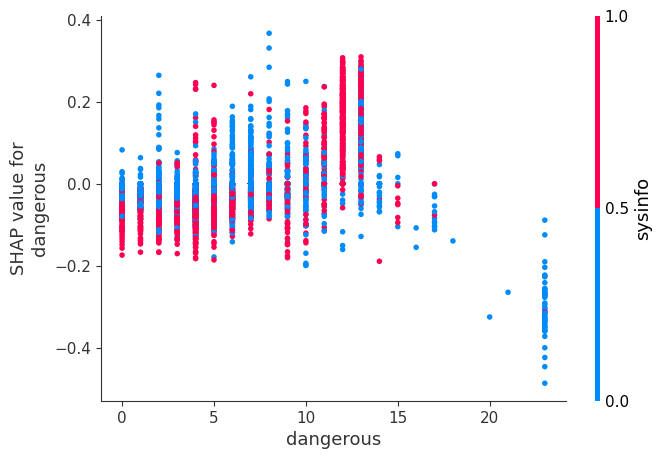}
  \\
  \adjustbox{scale=0.95}{(a) RBF-SVM}
  &
  \adjustbox{scale=0.95}{(b) $k$-NN}
  \\ \\[-1.5ex]
  \includegraphics[scale=0.35]{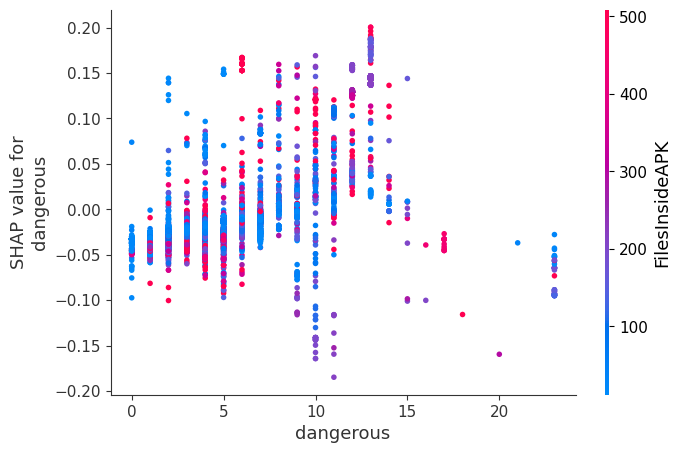}
  &
  \includegraphics[scale=0.35]{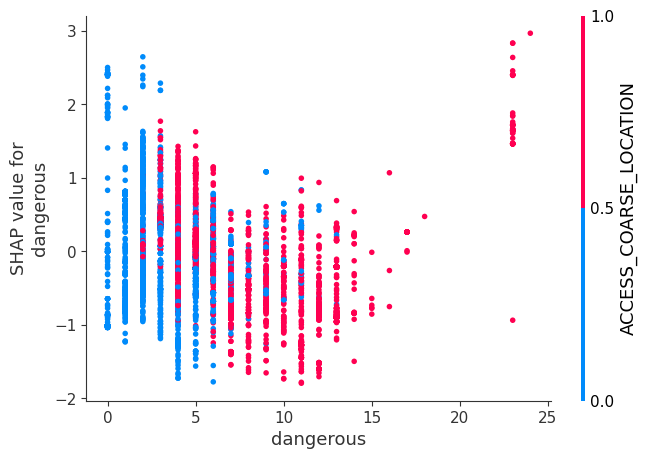}
  \\
  \adjustbox{scale=0.95}{(c) Random Forest}
  &
  \adjustbox{scale=0.95}{(d) MLP}
  \end{tabular}
  \caption{PDP plots (\texttt{dangerous})}\label{fig:pdp}
\end{figure}

We generate PDP plots using the \texttt{dependence\uu plot} method. This function 
automatically includes as the second variable the feature that interacts most strongly
with the selected variable. PDP plots with the \texttt{dangerous}
feature selected are shown in 
Figures~\ref{fig:pdp}(a) through~(d) 
for our RBF-SVM, $k$-NN, Random Forest, and MLP models, respectively. 
We note that the \texttt{dangerous} feature is discrete, with
values in the set~$\{0,1,2,\ldots,25\}$.

We observe that for the RBF-SVM model in Figure~\ref{fig:pdp}(a)
there is an approximately linear relationship between the raw
value of \texttt{dangerous} in the range from~0 to~13
and the corresponding Shapley values. Furthermore, over the 
range of~2 to~13, higher \texttt{dangerous} values are associated
with a progressively higher proportion of high values for 
\texttt{ACCESS\uu COARSE\uu LOCATION}, and 
beyond~13, only high values of \texttt{ACCESS\uu COARSE\uu LOCATION}
occur.

\begin{figure}[!htb]
  \centering
  \begin{tabular}{cc}
  \includegraphics[scale=0.35]{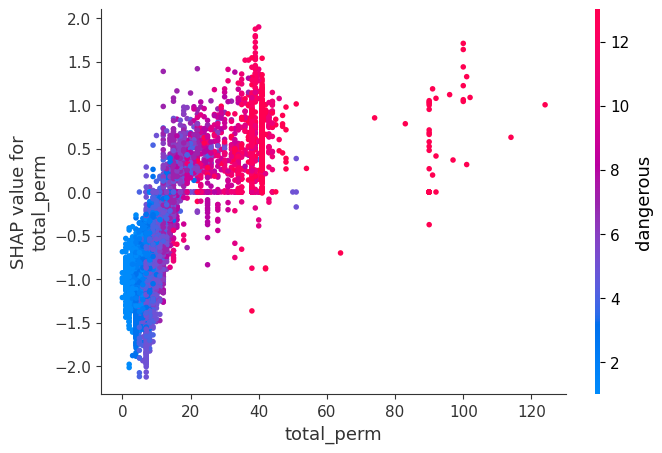}
  &
  \includegraphics[scale=0.35]{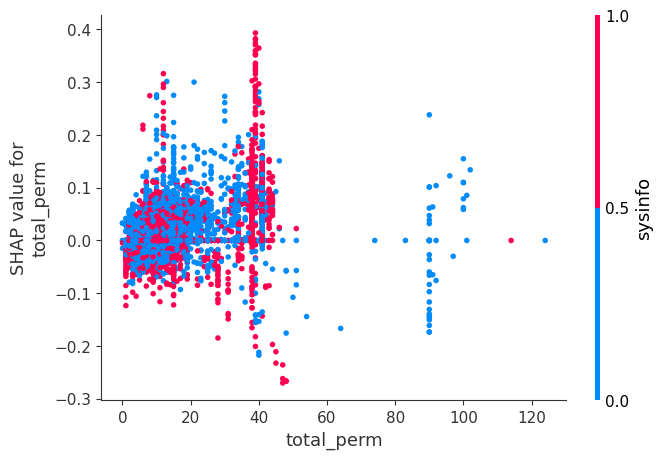}
  \\
  \adjustbox{scale=0.95}{(a) RBF-SVM}
  &
  \adjustbox{scale=0.95}{(b) $k$-NN}
  \\ \\[-1.5ex]
  \includegraphics[scale=0.35]{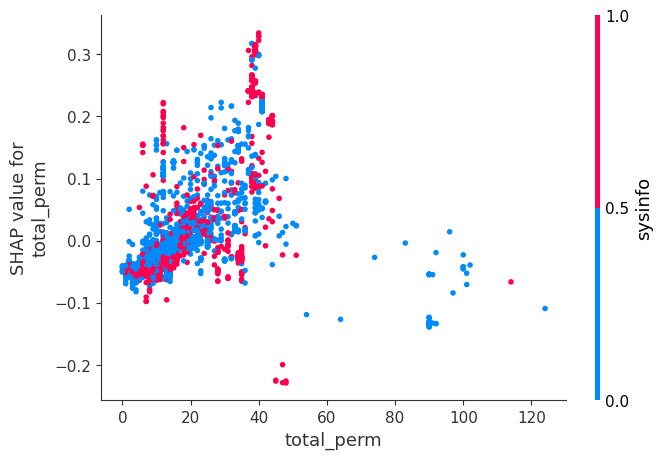}
  &
  \includegraphics[scale=0.35]{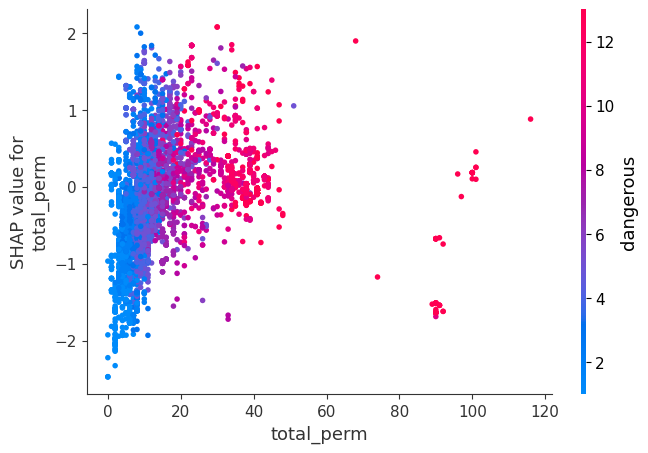}
  \\
  \adjustbox{scale=0.95}{(c) Random Forest}
  &
  \adjustbox{scale=0.95}{(d) MLP}
  \end{tabular}
  \caption{PDP plots (\texttt{total\uu perm})}\label{fig:pdp2}
\end{figure}

Figures~\ref{fig:pdp2}(a) through~(d) 
show PDP plots with the feature \texttt{total\uu perm} selected
for our RBF-SVM, $k$-NN, Random Forest, and MLP models, respectively.
We observe that the RBF-SVM model in Figure~\ref{fig:pdp2}(a)
shows a linear relationship between the raw value of the \texttt{total\uu perm}
and the Shapley values. Also, below a \texttt{total\uu perm} value
of about~10, the corresponding \texttt{dangerous} values are low,
while above that threshold, they are predominantly high.

Finally, we illustrate a local explanation for an individual sample using 
the SHAP \texttt{force\uu plot} method. 
This method requires the following three inputs.
\begin{enumerate}
\item The average of the model output over the training data, which
serves as the base value used to generate the force plot. 
\item The Shapley values, as computed on training data. 
\item The sample for which we wish to obtain a local explanation. 
\end{enumerate}

\begin{figure}[!htb]
\centering
\includegraphics[width=130mm]{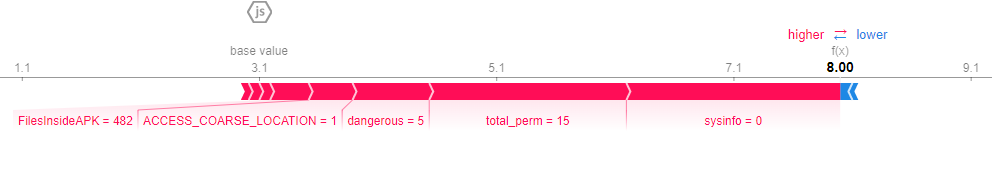}
\vglue -8pt
\caption{SHAP explanations for MLP (last observation)}
\label{fig:shaplocalmlp}
\end{figure}

Figure~\ref{fig:shaplocalmlp} shows the SHAP force plot generated 
for our MLP model, based on the last sample in the test dataset. 
Features that push the prediction higher (to the right) are shown in red, 
while those pushing the prediction lower are in blue.
In this case, the base value is~3.1, and based on the Shapley values,
\texttt{sysinfo} and \texttt{total\uu perm} 
have highest positive impact on the classification, with
\texttt{dangerous},
\texttt{ACCESS\uu COARSE\uu LOCATION}, and
\texttt{FilesInsideAPK} also having positive impact.
For this particular sample, no features have a significant negative
impact on the classification, as indicated by the lack of 
any ``force'' in the blue direction.

In summary, Shapley values indicate how much a feature 
contributes to the prediction of a given sample, and this
contribution can be positive or negative. If a feature is positively 
correlated to the target at a value higher than the average, it will contribute 
positively to the prediction. On the other hand, if a feature is negatively correlated 
to the target, it will contribute negatively to the prediction.
Furthermore, a wealth of information can be gleaned from Shapley
values using a number of different plotting strategies.



\section{Conclusion and Future Work}\label{chap:conclusion}

In this paper, we provided a selective survey of previous work involving the 
application of XAI techniques to detection and classification problems 
in the malware domain. We then
performed a comparative study of several XAI techniques 
for a variety of models, including classic ML models 
(linear SVM, RBF-SVM, Random Forest, and $k$-NN) 
and deep learning models (MLP and CNN). When trained on 
a challenging Android malware multiclass 
problem, we found that Random Forest performed best among these models,
followed closely by MLP, with all of the models performing within a few percentage points
of the best model.

We applied a several well-known XAI techniques (ELI5, LIME, CAM, and SHAP) 
to our trained models. All of these XAI techniques provided interesting information
about the learning models to which they were applicable. 
Although relatively costly to compute, SHAP explanations were
particularly informative. 

ELI5 proved effective at providing global explanations, while 
LIME generated explanations at a granular level of individual samples.
CAM uncovered details of the inner workings of our CNN 
model, which otherwise would have remained very opaque.
SHAP provided many insights, including PDP plots 
that enabled us to visualize relationships between pairs of features.



There are many potential avenues for future research. 
It would certainly be useful to have guidelines for 
determining which XAI techniques are most likely to produce
useful results for problems in the malware domain. Of course, it would also
be useful to have such guidelines more generally, that is,
for a given model type when trained on a dataset
from a specific problem domain. Additional work
to quantify XAI results is another important fundamental research topic. 

Finally, we note that the work in~\cite{2020DBLP} purports to show that 
``$\ldots$ explanation results obtained in the malware analysis domain
cannot achieve a consensus in general $\ldots$''. 
Some of our results presented in Section~\ref{chap:results} 
do raise questions of consistency. This issue of consistency
(or lack thereof) is perhaps the most pressing concern
in the entire field of XAI, and hence further research 
on this topic is needed.

\bibliographystyle{plain}
\bibliography{references.bib}

\clearpage


\section*{Appendix}
\renewcommand{\thesubsection}{A.\arabic{subsection}}
\setcounter{table}{0}
\renewcommand{\thetable}{A.\arabic{table}}
\setcounter{figure}{0}
\renewcommand{\thefigure}{A.\arabic{figure}}

\begin{figure}[!htb]
\centering
\begin{tabular}{|c|c|c|c|} \hline
\rotatebox{90}{\includegraphics[scale=0.425]{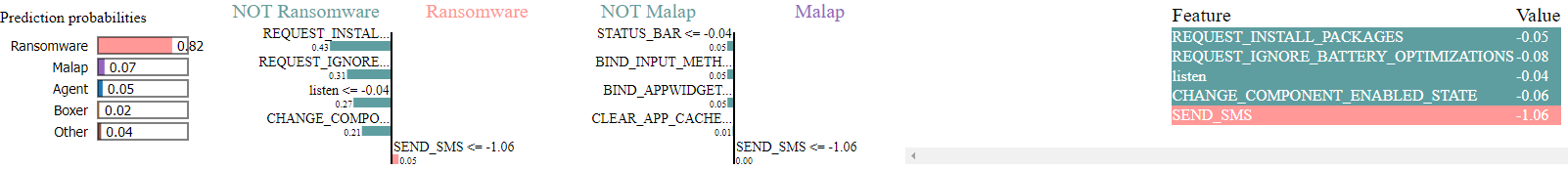}} 
&
\rotatebox{90}{\includegraphics[scale=0.425]{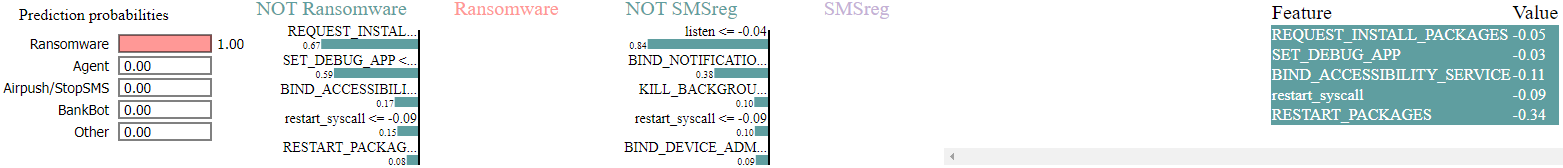}} 
&
\rotatebox{90}{\includegraphics[scale=0.425]{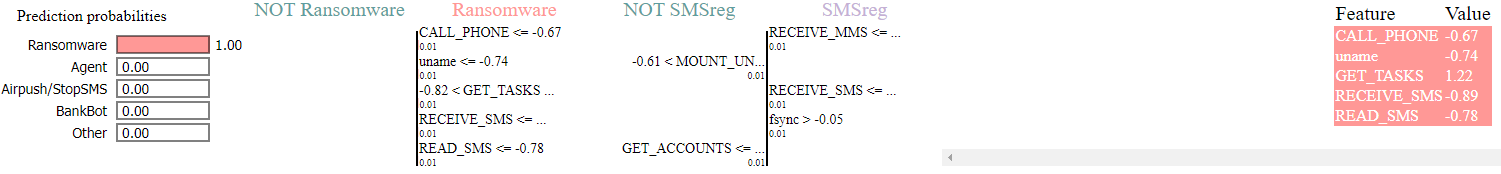}} 
&
\rotatebox{90}{\includegraphics[scale=0.425]{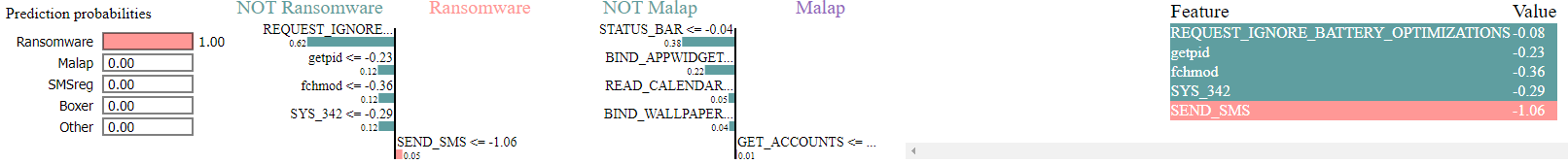}} 
\\ \hline
\multicolumn{1}{c}{\raisebox{-0.75ex}{\adjustbox{scale=0.95}{(a) RBF-SVM}}} 
&
\multicolumn{1}{c}{\raisebox{-0.75ex}{\adjustbox{scale=0.95}{(b) $k$-NN}}}
&
\multicolumn{1}{c}{\raisebox{-0.75ex}{\adjustbox{scale=0.95}{(c) Random Forest}}}
&
\multicolumn{1}{c}{\raisebox{-0.75ex}{\adjustbox{scale=0.95}{(d) MLP}}}
\end{tabular}
\caption{LIME explanations for correctly classified sample}\label{fig:lime_correct}
\end{figure}

\begin{figure}[!htb]
\centering
\begin{tabular}{|c|c|c|c|} \hline
\rotatebox{90}{\includegraphics[scale=0.425]{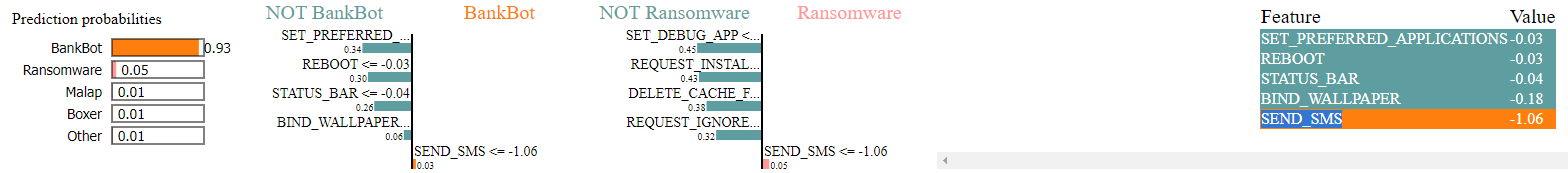}} 
&
\rotatebox{90}{\includegraphics[scale=0.425]{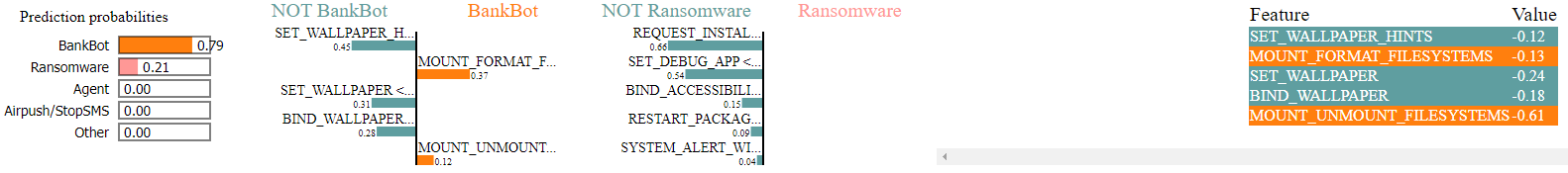}} 
&
\rotatebox{90}{\includegraphics[scale=0.425]{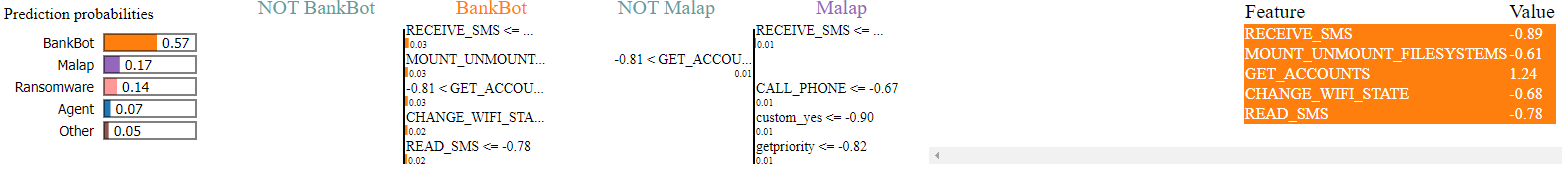}} 
&
\rotatebox{90}{\includegraphics[scale=0.425]{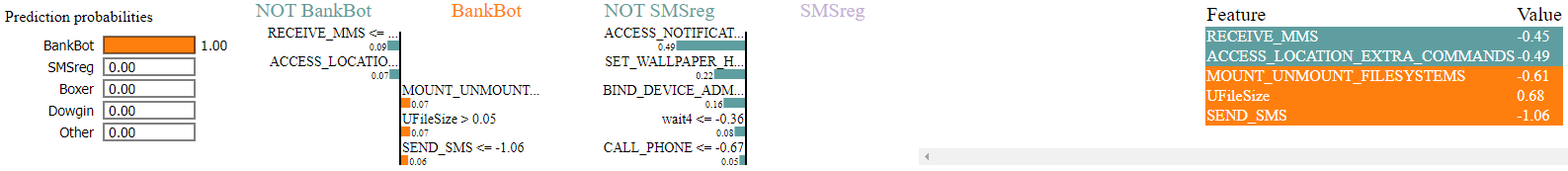}} 
\\ \hline
\multicolumn{1}{c}{\raisebox{-0.75ex}{\adjustbox{scale=0.95}{(a) RBF-SVM}}} 
&
\multicolumn{1}{c}{\raisebox{-0.75ex}{\adjustbox{scale=0.95}{(b) $k$-NN}}}
&
\multicolumn{1}{c}{\raisebox{-0.75ex}{\adjustbox{scale=0.95}{(c) Random Forest}}}
&
\multicolumn{1}{c}{\raisebox{-0.75ex}{\adjustbox{scale=0.95}{(d) MLP}}}
\end{tabular}
\caption{LIME explanations for incorrectly classified sample}\label{fig:lime_incorrect}
\end{figure}

\end{document}